\documentclass[12pt]{iopart}

\usepackage{graphicx}
\usepackage{color}
\usepackage{bm}
\usepackage{ulem}
\usepackage{CJK}

\usepackage{amssymb}
\usepackage{cite}
\usepackage[colorlinks,linkcolor=blue,anchorcolor=blue,citecolor=blue,urlcolor=blue]{hyperref}
\begin{document}
\title[Electron and ion acceleration from femtosecond laser-plasma peeler scheme]{Electron and ion acceleration from femtosecond laser-plasma peeler scheme}

\author{X. F. Shen$^{1,\ast}$, A. Pukhov$^{1,\ast}$, B. Qiao$^{2}$}
\address{$^1$ Institut f\"ur Theoretische Physik I, Heinrich-Heine-Universit\"at D\"usseldorf,
	40225 D\"usseldorf, Germany}
\address{$^2$ Center for Applied Physics and Technology, HEDPS, SKLNP, and School of Physics, Peking University, Beijing, 100871, China}
\ead{shenx@uni-duesseldorf.de and pukhov@tp1.uni-duesseldorf.de}
\vspace{10pt}

\date{\today}

\begin{abstract}

Using three-dimensional particle-in-cell simulations, we further investigate the electron and ion acceleration from femtosecond laser-plasma peeler scheme which was proposed in our recent paper (Shen \textit{et al} 2021 \textit{Phys. Rev. X} \textbf{11} 041002). In addition to the standard setup where a laser pulse impinges on an edge of a single tape target, two new variants of the target, i.e., a parallel tape and a cross tape target, were proposed, where strong surface plasma waves can also be efficiently excited at the front edges of the target. By using a tabletop 200 TW-class laser pulse, we observe generation of high-flux, well-collimated, superponderomotive electrons. More importantly, quasimonoenergetic proton beams can always be obtained in all the three setups, while  with the single tape case, the obtained proton beam has the highest peak energy and narrowest spectrum.

\end{abstract}

\maketitle

\section{Introduction}

Laser-driven ion acceleration \cite{Macchi2013,Daido2012,Macchi2017} has a wide range of prospective applications including proton radiography for ultrafast science and implosion dynamics \cite{Borghesi2001,Li2006}, production of warm dense matter  \cite{Brambrink2007,Schollmeier2008}, nuclear physics \cite{Beg2002}, tumor therapy \cite{Bulanov2002c}, fast ignition \cite{Roth2001}, injectors for conventional accelerators \cite{Noda2006}, etc. Many of these applications prefer quasimonoenergetic ion beams. For ion beam therapy, the required energy spread is only about 1$\%$. Even the ideal energy spectrum, to deposit the so-called ``spread-out Bragg peak" (SOBP) dose distribution, actually has a prominent peak at the cutoff energy, instead of a decaying profile \cite{Bulanov2002c}. 
In ion fast ignition, ion beams with monoenergetic spectrum have several advantages over those with broad spectrum, such as their better coupling with compressed fuel (which means lower ignition threshold) and the possibility to place the ion source farther from the fuel without using re-entrant cones (reducing the difficulties in target fabrication) \cite{Hegelich2011,Honrubia2009}.  
Moreover, for production of warm dense matter, proton beams with narrow spectrum can help achieve uniform heating, rather than volumetric heating from broad spectrum \cite{Schollmeier2008}.

Thanks to the advancements in both laser technology and targetry, recently ion beam qualities and shot-to-shot reproducibility have been significantly improved. Proton beams with cutoff energy near 100 MeV \cite{Higginson2018,Wagner2016,Kim2016} and carbon ions of about 50 MeV/u \cite{Ma2019} have been demonstrated independently in several laser systems. Tens of consecutive shots were conducted to evaluate the stability and reproducibility of laser-ion acceleration \cite{Steinke2020,Ziegler2021}. Last year, pilot experiments about laser-based ion beam therapy were reported \cite{Bin2022,Kroll2022}. The single shot doses exceed 20 Gy over millimeter-scale volumes and the instantaneous doses are up to 10$^9\,{\rm Gy/s}$. This enables the possibilities for the study of ultrahigh dose-rate FLASH radiotherapy \cite{Favaudon2014}.  However, currently the obtained ion energy spectra are usually exponentially decaying and the maximum energy is relatively low compared to the requirements of many applications \cite{Macchi2013,Daido2012,Macchi2017,Higginson2018,Steinke2020,Ziegler2021}. Though a sophisticated energy-selection system can be used to select ions out of a broad energy spectrum, it not only causes huge particle losses, but also would significantly increase the system size, especially considering that a gantry that contains a flexible patient couch and a rotating
beam line is necessary for future clinical treatment \cite{Linz2016}.

Different ion acceleration mechanisms have been proposed in the past two decades, including target normal sheath acceleration (TNSA) \cite{Schreiber2006,Snavely2000,Pukhov2001,Fuchs2006,Robson2007,Mora2003,Kiefer2013,Nakatsutsumi2018}, radiation pressure acceleration (RPA) \cite{Qiao2009,Yan2008,Robinson2008,Qiao2010,Esirkepov2004,Macchi2009,Shen2017,Shen2021a,Henig2009,Kar2012,Weng2012,Scullion2017}, collisionless shockwave acceleration (CSA) \cite{Zhang2016,Fiuza2012,Haberberger2012,Liu2016}, magnetic vortex acceleration (MVA) \cite{Bulanov2005,Nakamura2010,Zhang2017a,Park2019,Reichwein2021}, Coulomb explosion \cite{Fourkal2005,Bulanov2002,Braenzel2015,Hilz2018}, ion wakefield acceleration \cite{Shen2007,Zheng2012,Liu2018}, etc. Each of these mechanisms can accelerate ions independently at certain range of laser and plasma parameters, but often, a hybrid acceleration that includes two or several of them occurs, such as the widely discussed hybrid RPA-TNSA \cite{Qiao2012,Higginson2018}, breakout afterburner acceleration (or relativistically-induced transparency) \cite{Yin2006}, hybrid CSA-TNSA \cite{Fiuza2012}, etc. This is because in realistic laser-plasma interactions, the light pressure carried by photons, thermal pressure carried by hot electrons and electrostatic pressure induced by charge separation always exist simultaneously \cite{Qiao2019}. Moreover, enhanced TNSA has been reported in various schemes, including but not limited to using cone-like targets \cite{Gaillard2011,Zou2017,Yang2018,Gizzi2020}, thin foil coated behind a long near-critical-density (NCD) plasma \cite{Bin2015,Guenther2022,Rosmej2020}, thin foil attached microwire arrays in front \cite{Wang2018,Jiang2016,Curtis2021} and multi-picosecond laser pulses \cite{Raffestin2021}. Cascaded ion acceleration has also been proposed in several different setups \cite{He2019,Gonoskov2009,Wang2019}.

To achieve monoenergetic ion beams, in the source part, there are two main methods. One is to utilize a longitudinal bunching and accelerating field under which fast ions experience a smaller field and the slow ones a larger field. This method is analogous to that used in conventional radio-frequency accelerators, where such a bunching field is realized by controlling the phase of the synchronous
particle relative to the crest of the accelerating wave \cite{Wangler1998}. In laser-ion acceleration, such a bunching field has been observed in the RPA \cite{Qiao2009,Yan2008,Macchi2009,Robinson2008} and CSA \cite{Fiuza2012}. However, RPA is plagued by strong electron heating due to effects of transverse instabilities \cite{Pegoraro2007} and finite spot size \cite{Dollar2012}. These effects may induce relativistic transparency and destroy the bunching field. As a consequence, the obtained ion energy is rather limited and the energy spread is large. Various approaches to suppress the instability development have been proposed \cite{Shen2017,Shen2017b,Bin2015,Yu2010,Bulanov2015,Ju2021,Tamburini2010,Chen2009}. For example, a scheme of dynamic-ionization stabilized RPA was proposed by Shen {\it et al.} in 2017 \cite{Shen2017} and demonstrated in very recent experiments \cite{Alejo2022}. Moreover, the CSA requires some special density profiles with peak density close to the critical density to launch a fast shock and suppress the TNSA \cite{Fiuza2012}. Such targets are not easy to fabricate for the short-wavelength powerful Ti:Sa or Nd:glass laser pulses. Only few experiments have successfully demonstrated the CSA \cite{Haberberger2012,Zhang2016}.

The other method is to accelerate only a small population of protons that are initially  distributed in a compact region such that they feel almost the same accelerating field. This can be achieved by putting a small dot behind a thick target \cite{Schwoerer2006}, using multi-species \cite{Shen2019} or multi-layer \cite{Bulanov2002} targets. The underlying mechanism could be TNSA, Coulomb explosion or other schemes. Experiments have demonstrated the effectiveness of this method \cite{Schwoerer2006,Hilz2018}. 
Nevertheless, until now, in both theory and experiment, how to produce high-energy (like
100-MeV protons), monoenergetic (energy spread of about 1$\%$) ion beams is still an open question.

Recently, we proposed a novel ion acceleration scheme named peeler scheme that can produce quasimonoenergetic proton beams with peak energy above 100 MeV, energy spread of about 1$\%$ and particle number of about $10^9$ by irradiating a petawatt femtosecond laser pulse on a tape target \cite{Shen2021b}. In this scheme, a longitudinal bunching and transverse focusing field for proton acceleration is self-established because a huge amount of electrons are peeled off at the front edge, accelerated to superponderomotive energies along the target lateral surfaces and injected into the target rear side. The obtained proton energy is higher than that from the standard TNSA and RPA. Moreover, a fresh paper by Sarma {\it et al.} \cite{Sarma2022} further investigated the electron and proton acceleration in this peeler scheme. 
More related discussions can also be found in very recent works \cite{Marini2022,Zhu2022}.

In this paper, we further discuss the experimental feasibility of this peeler scheme by proposing two variants that may help identify its effectiveness and key feature (i.e., narrow spectrum) in experiments. The schematics are shown in Fig. \ref{fig:fig1}. Besides the standard single tape target [Fig. \ref{fig:fig1}(a)] discussed in Ref. \cite{Shen2021b}, we propose to use two alternatives, i.e., a target contains several parallel tapes [Fig. \ref{fig:fig1}(b), hereinafter referred to the parallel tape case] or a cross tape [Fig. \ref{fig:fig1}(c), hereinafter referred to the cross tape case]. Such targets may help mitigate the effects of the pointing stability that exists in most of current laser facilities. Note that the pointing stability of the state-of-the-art laser facilities is about 1$\mu$rad \cite{Nakamura2017,Danson2015} that is good enough to ensure that in most shots, the center of the laser pulse can impinge on the tape. It can be further reduced by implementing adaptive focusing optics \cite{Ohland2022}. Moreover, we mention that from the long-term view, a small pointing stability is necessary for most laser systems to achieve stable and reproducible operation, not only for our scheme, but also for most other experiments such as laser-driven particle acceleration with structural (or mass-limited) targets \cite{Hilz2018,Obst2017,Zhang2017a}, collision of high-energy electrons with intense laser pulses \cite{Cole2018,Poder2018} and laser-plasma based x-ray free electron laser \cite{Wang2021}. 

This paper is organized as follows. In section II, we introduce the simulation setups and parameters. In section III, we display the main simulation results, including electron acceleration, forward proton and carbon ion acceleration, lateral gold ion acceleration, and compare the results from different setups. In section IV, we discuss the differences between 2D and 3D simulations and the robustness of our scheme. Conclusion are given in sections V.

\section{Simulation setup}

\begin{figure*}[t]
	\includegraphics[width=14cm]{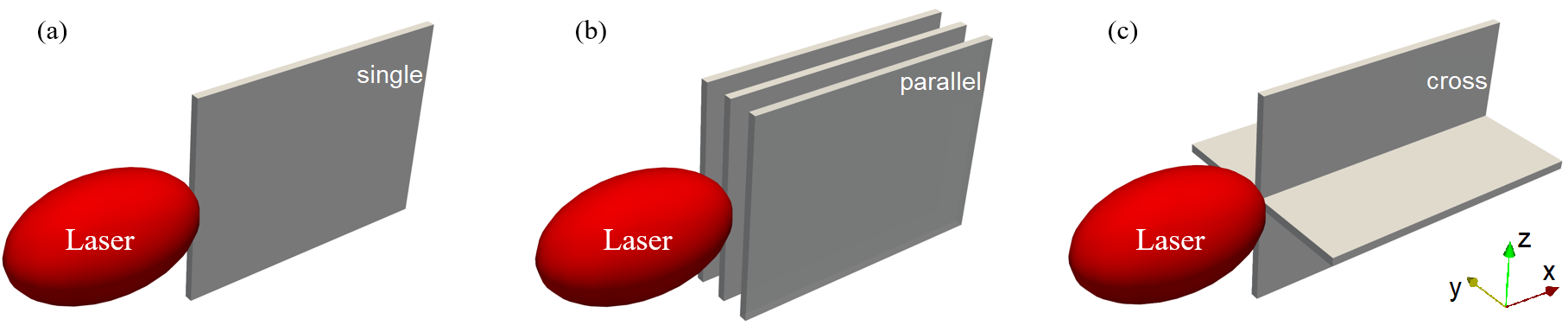} \caption{Schematics of the simulation setups: a femtosecond intense laser pulse impinges on a single tape target (a), a parallel tape target (b) and a cross tape target. Note that in experiment, the target size along $z$-direction can be infinitely long and for the setups in (b) and (c), more parallel tapes or periodic-distributed crosses (like mesh targets) can be used.
	}
	\label{fig:fig1} 
\end{figure*}

Three-dimensional (3D) particle-in-cell (PIC) simulations are conducted with the \textsc{EPOCH} code \cite{Arber2015}. Different from our previous work \cite{Shen2021b}, here we focus on comparing the results from the variant setups shown in Fig. \ref{fig:fig1} to demonstrate that quasimonoenergetic proton beams can be stably obtained in the peeler scheme, instead of optimizing the parameters to achieve extremely narrow spectrum. Hence, we consider a tightly-focused laser pulse with intensity of $I_0=8.65\times10^{20}\,{\rm W/cm^2}$ ($a_0=20$), wavelength of $\lambda=800$ nm, focal spot size of $d_L=6\lambda$ (full-width at half-maximum, FWHM in intensity) and pulse duration of $\tau_L=25$ fs (FWHM in intensity). The spatial and temporal profiles are both Gaussian distributions.  The corresponding laser power is about 240 TW. Such laser systems have been widely constructed around the world \cite{Danson2015}. 
A plasma tape of high-$Z$ material (we assume gold) has dimensions $x\times y\times z=30\lambda\times0.75\lambda\times23\lambda$. 
To save  the computational resources, the electron densities for the plasma tape and CH layer are both chosen as relativistically overdense $n_e=30n_c$ where $n_c=\pi m_ec^2/e^2\lambda^2$ is the critical density, $c$ is the light speed, $m_e$ and $e$ are the electron mass and charge. The initial charge states of ions are given according to the Ammosov-Delone-Krainov formula \cite{Perelomov1966}, which means the ions species are ${\rm Au}^{51+}$ in the plasma tape, and ${\rm C^{6+}}$ and ${\rm H}^{+}$ in the hydrocarbon layer. The longitudinal thickness of the CH layer is set to $0.4\lambda$ to ensure a sufficiently large proton areal density for facilitating our physical discussion and mimicking the components of the contaminants, and the transverse sizes are the same as those of the plasma tape. The density ratio of proton to carbon ions is $n_{\rm H^{+}}:n_{\rm C^{6+}}=1:1$. 
The simulation box is $70\lambda\times24\lambda\times24\lambda$ in the $x\times y\times z$ directions, containing $2800\times960\times480$ cells, respectively. Here, a higher resolution in
the $y$-direction is used to resolve the plasma skin depth and the process of extracting electrons from the tape since a $y$-polarized laser pulse is considered. 
The macroparticles in each cell for electrons, ${\rm Au}^{51+}$, ${\rm C^{6+}}$ and ${\rm H}^{+}$ are 8, 1, 8 and 32, respectively. Open boundary conditions are used for particles and fields in each direction. 
In the parallel tape case, to reduce computational cost, we only consider three identical tapes that are parallel to each other and the gap inbetween is set to $d_L/2$, see Fig. \ref{fig:fig1}(b).  
In the cross tape case, the density is uniform and the size of the horizontal tape along $y$-direction is $23\lambda$, see Fig. \ref{fig:fig1}(c).

\section{Simulation results}

\subsection{Results from a single tape target}

\begin{figure*}[t]
	\includegraphics[width=16cm]{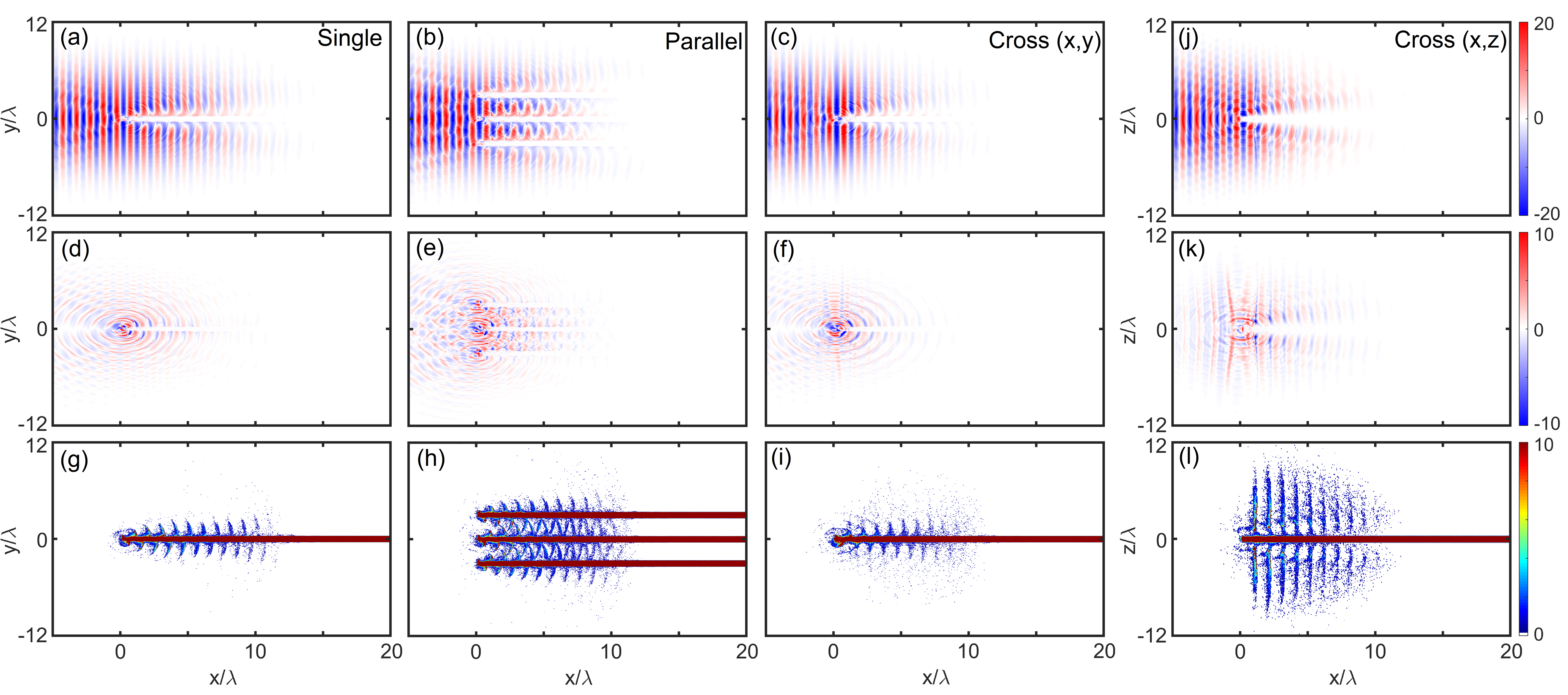} \caption{Comparison of the transverse field $E_y$ [(a)-(c)], longitudinal field $E_x$ [(d)-(f)] and electron density $n_e$ [(g)-(i)] from the single [(a),(d),(g)], parallel [(b),(e),(h)] and cross [(c),(f),(i)] tape cases at $t=-1T_0$. For the cases with a single or parallel tape target, the 2D maps are cut in the $z=0$ planes while for the cross tape case, they corresponds to the slices in the $z=0.9\mu$m plane. (j)-(l) show the corresponding distributions from the cross tape case in the  $y=0.9\mu$m plane. Here the electric fields are normalized to $m_ec\omega_0/e$ and the density is normalized to the critical density $n_c$.
	}
	\label{fig:fig2} 
\end{figure*}

Figure \ref{fig:fig2} displays the simulation results, where the transverse field $E_y$, longitudinal field $E_x$ and electron density $n_e$ at $t=-1T_0$ from the single tape case are shown in (a), (d) and (g), respectively. Here $t=0$ corresponds to the time when the laser peak impinges on the front edge of the target. One can clearly see that though the pulse duration and the spot size used here are quite different from our previous work \cite{Shen2021b}, the main physical phenomena are similar, i.e., a strong surface plasma wave (SPW) is efficiently excited at the front edge of the target [Fig. \ref{fig:fig2}(d)], electrons are periodically peeled off at the edge and accelerated forward along the target lateral surfaces [Fig. \ref{fig:fig2}(g)]. The amplitude of the $E_x$ field exceeds $4\times10^{13}\,{\rm V/m}$. The electron nanobunches with high density are well-collimated [Fig. \ref{fig:fig6}(a)] and will be accelerated to superponderomotive energies [red line in Fig. \ref{fig:fig7}(a)] by both the longitudinal and transverse fields \cite{Shen2021b,Wang2018}. At $t=29T_0$, the highest electron energy is about 250 MeV and the effective electron temperature $T_{\rm eff}$ is about 35 MeV, both comparable to the results shown in Ref. \cite{Shen2021b}. The number of electrons with $\gamma>10$ reaches about 41.4 nC.

\begin{figure*}[t]
	\includegraphics[width=14cm]{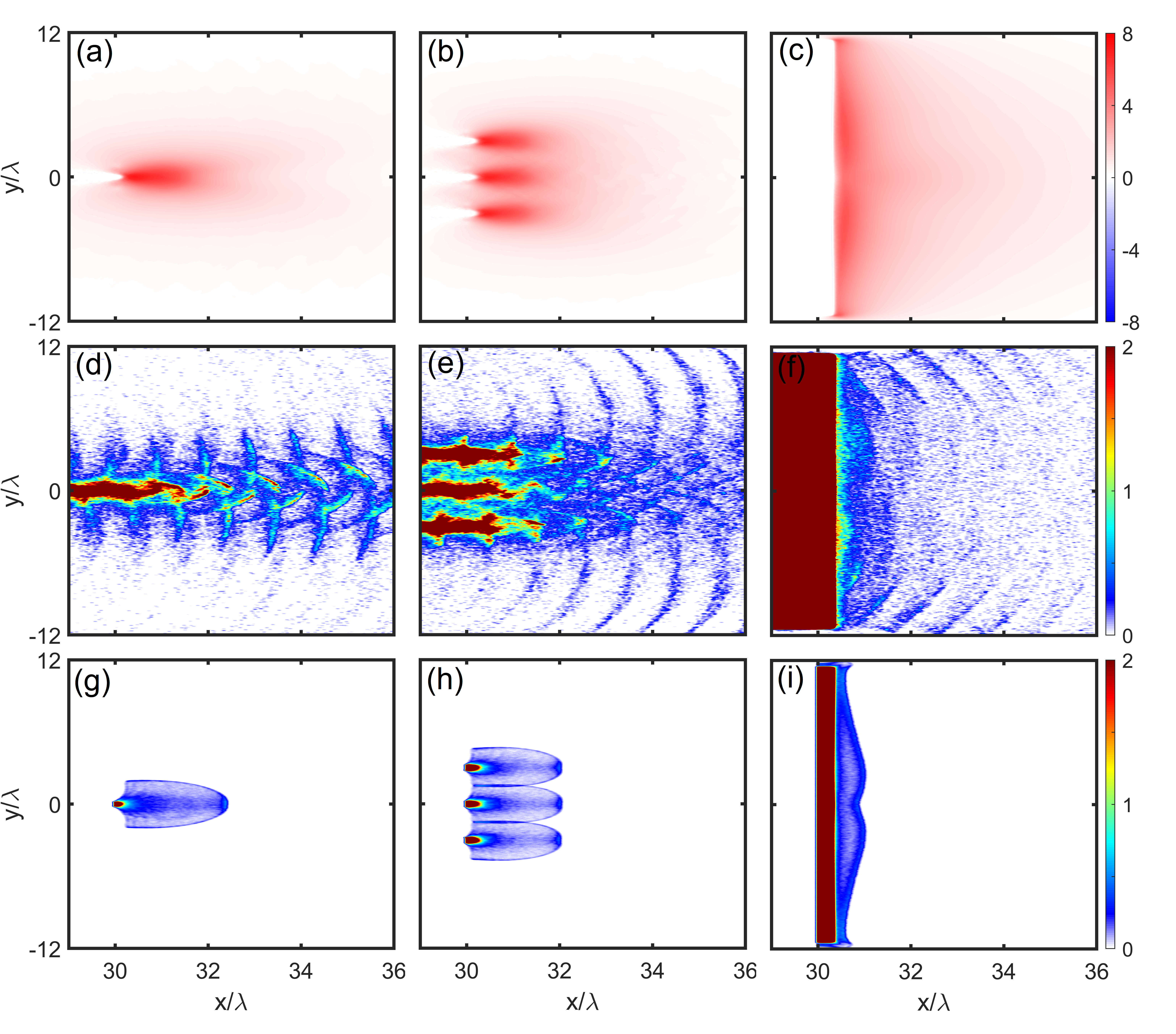} \caption{Comparison of the longitudinal field $E_x$ [(a)-(c)], electron density $n_e$ [(d)-(f)] and proton density $n_p$ [(g)-(i)] from the single [(a),(d),(g)], parallel [(b),(e),(h)] and cross [(c),(f),(i)] tape cases at $t=34T_0$ in the $z=0$ plane. Note that here the $E_x$ field is time-averaged to show the quasistatic components, same for Fig. \ref{fig:fig4}.
	}
	\label{fig:fig3} 
\end{figure*}

\begin{figure*}[t]
	\includegraphics[width=14cm]{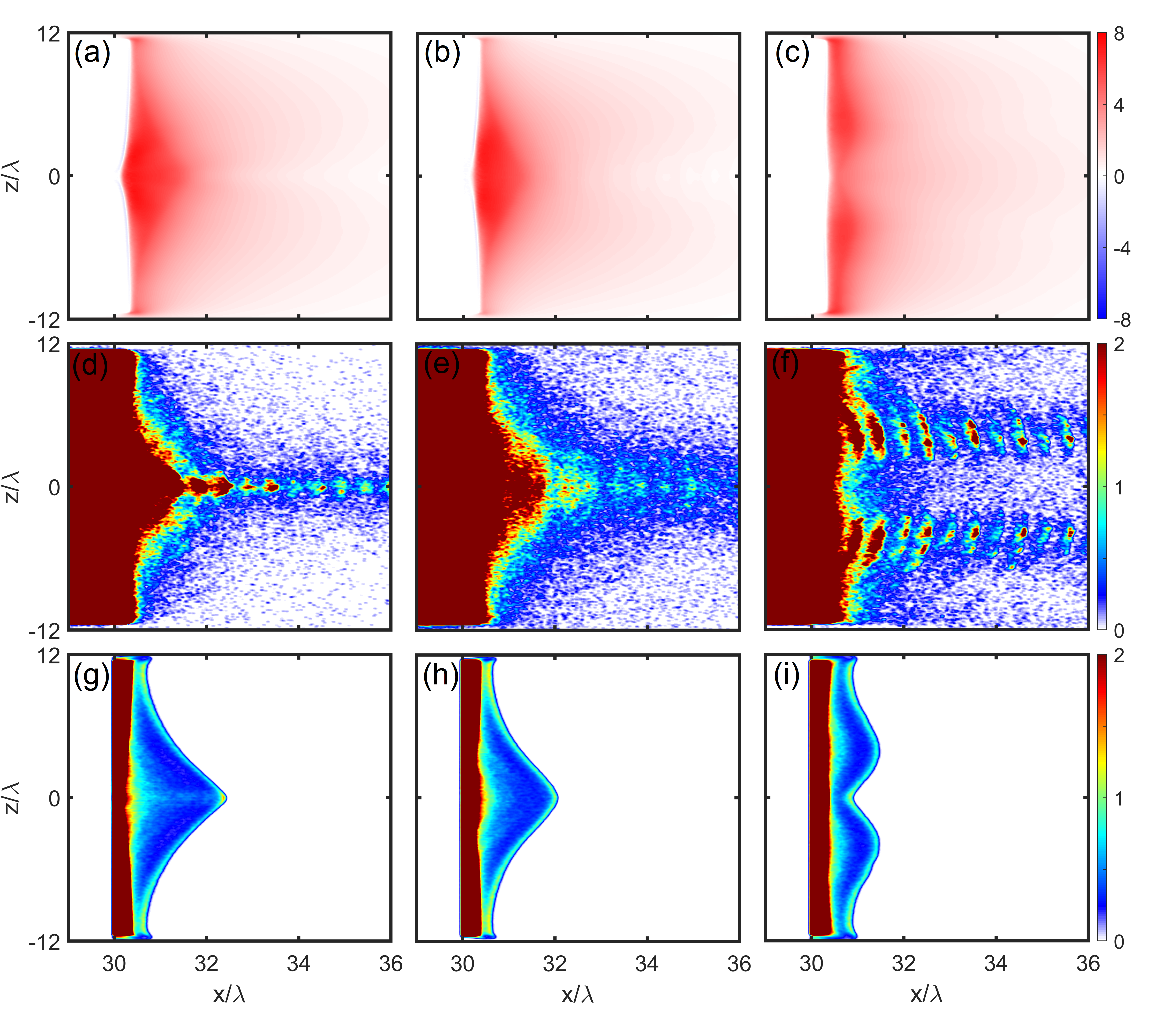} \caption{Comparison of the longitudinal field $E_x$ [(a)-(c)], electron density $n_e$ [(d)-(f)] and proton density $n_p$ [(g)-(i)] from the single [(a),(d),(g)], parallel [(b),(e),(h)] and cross [(c),(f),(i)] tape case  at $t=34T_0$ in the $y=0$ plane.
	}
	\label{fig:fig4} 
\end{figure*}

\begin{figure*}[h]
	\includegraphics[width=14cm]{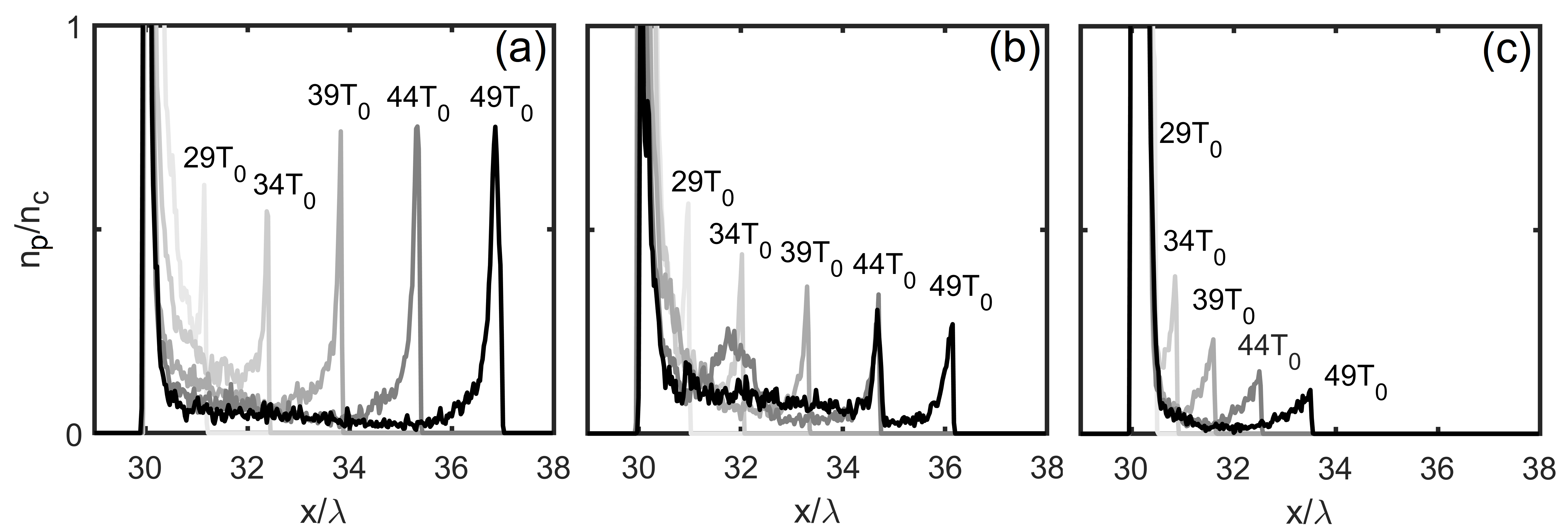} \caption{Temporal evolution of the proton density distribution on the $x$-axis for the case with a single (a), parallel (b) and cross (c) tape target.
	}
	\label{fig:fig5} 
\end{figure*}

\begin{figure*}[t]
	\includegraphics[width=14cm]{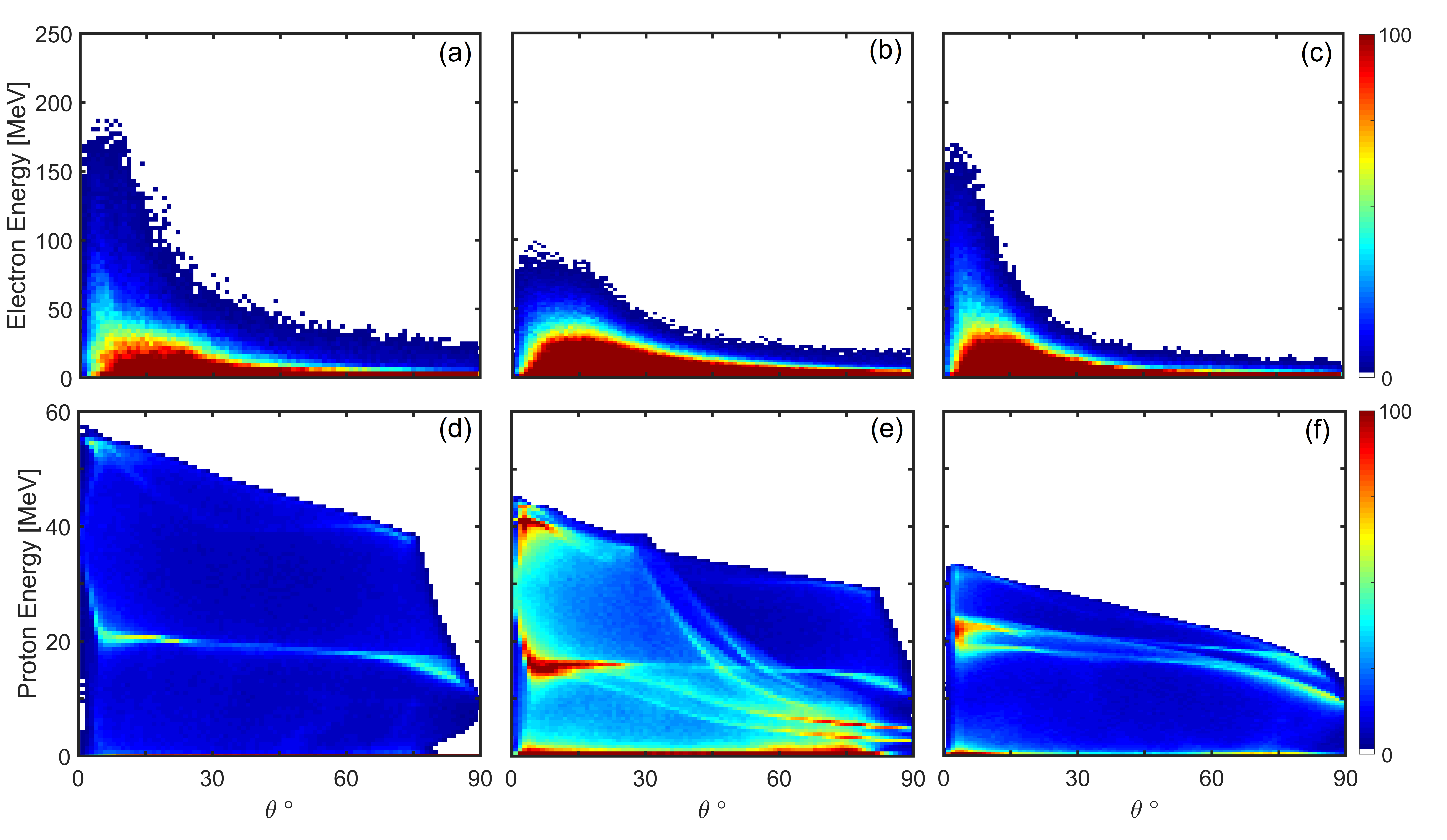} \caption{Comparison of angular distributions of electrons [(a)-(c)] at $t=29T_0$ and protons [(d)-(f)] at $54T_0$ from a single [(a),(d)], parallel [(b),(e)] and cross [(c),(f)] tape target. Note that here the number of particles is normalized to the same value in different cases. The second peaks seen in the proton angular distributions come from the finite simulation box in $z$-direction, as detailed explained in the text.
	}
	\label{fig:fig6} 
\end{figure*}

When these high-density electron nanobunches are injected into the rear vacuum [see Figs. \ref{fig:fig3}(d) and \ref{fig:fig4}(d)], a strong quasistatic longitudinal field is induced there, as shown in Figs. \ref{fig:fig3}(a) and \ref{fig:fig4}(a). Its amplitude reaches about $3\times10^{13}\,{\rm V/m}$ at $t=34T_0$, leading to efficient ion acceleration. This is a bunching field for proton acceleration.  Under this field, fast protons are accelerated by a smaller $E_x$ field while slower protons by a larger $E_x$. Therefore, the slower ones may catch up with the faster, forming a density peak, see  Figs. \ref{fig:fig3}(g) and \ref{fig:fig4}(g). In Fig. \ref{fig:fig5}(a), we show the 1D cut of the proton density along $x$-axis, where a prominent proton density peak appears and exists for a long time. The final angular distribution and energy spectrum of the protons are shown in Fig. \ref{fig:fig6}(d) and the red line in Fig. \ref{fig:fig7}(b), respectively. In Fig. \ref{fig:fig6}(d), one can see that there is an isolated well-collimated peak at the highest energy range. From the red line in Fig. \ref{fig:fig7}(b), we know that the maximum proton energy is about 57.3 MeV and the peak energy is about 54.7 MeV. The energy spread is about 3$\%$ and there are about $2\times10^8$ protons within the peak (FWHM). The total conversion efficiency from laser to protons is about 0.5$\%$. Though it is not very high, a large part of the energy are carried by the high-energy protons, unlike that in a decaying spectrum obtained from other mechanisms.

Note that here the results are not as good as those we shown in Ref. \cite{Shen2021b} since we consider different laser and plasma parameters here, especially that the pulse duration is much shorter, leading to shorter (fewer) electron nanobunches that can be used in proton acceleration. However, this does not mean the longer the laser pulse duration, the better the electron and proton acceleration, since one has to consider the transverse expansion of the target. The expanding plasma may block the following laser pulse. Hence it exists an optimal pulse length by considering the target expansion. Moreover, the lower peak around 20 MeV is caused by the finite target size along $z$-direction. As shown in Fig. \ref{fig:fig4}(g), since electrons can be pulled back transversely at the side edges at which the longitudinal field is relatively strong and protons are being accelerated. If a longer target along $z$-direction is used, this peak can be strongly suppressed, see Ref. \cite{Shen2021b} and also the 2D simulation results shown below.

\subsection{Results from a parallel tape target}

\begin{figure*}[h]
	\includegraphics[width=14cm]{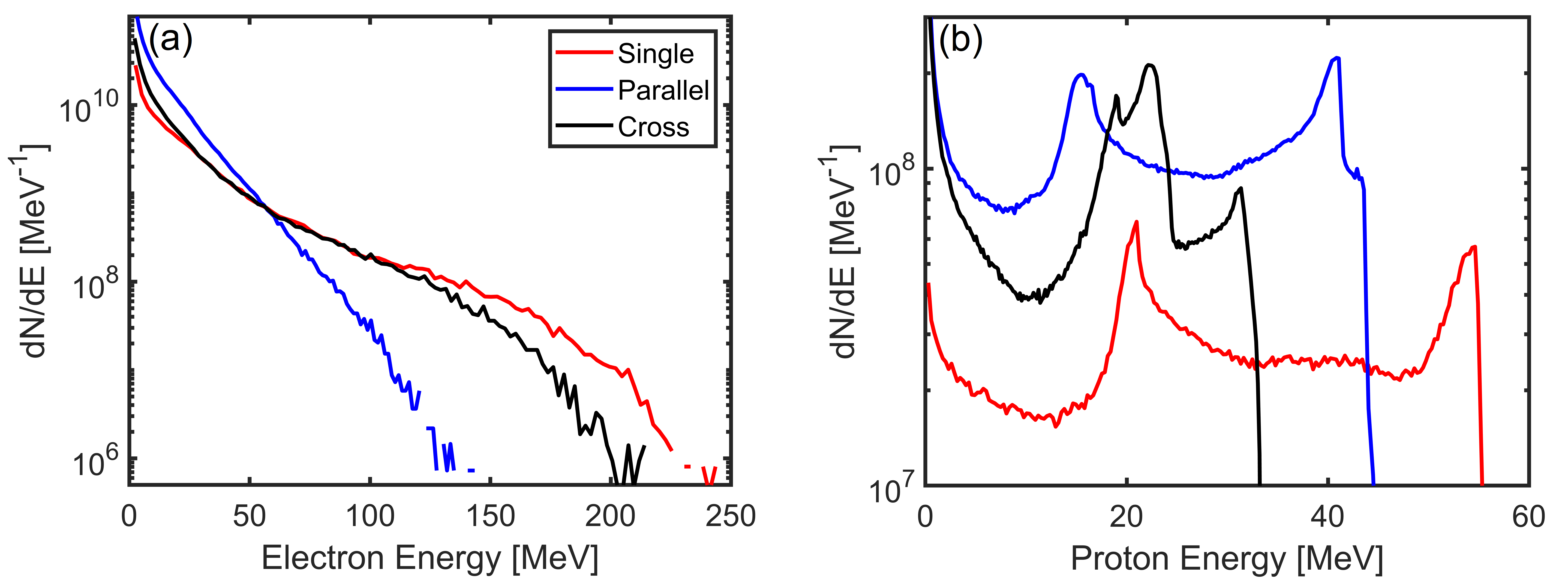} \caption{Comparison of electron (a) and proton (b) acceleration from the single (red), parallel (blue) and cross (black) tape target. For electrons, we only consider those forward-moving ones at $t=29T_0$ when the laser peak reaches the rear edge of the target and the effective electron temperature is the highest. Protons shown in (b) are selected with $|\theta|<10^\circ$ at $t=54T_0$ where $\theta={\rm atan}(\sqrt{p_y^2+p_z^2}/p_x)$.
	}
	\label{fig:fig7} 
\end{figure*}

For the parallel tape case, Figs. \ref{fig:fig2}(b), \ref{fig:fig2}(e) and \ref{fig:fig2}(h) show the distributions of the transverse field $E_y$, longitudinal field $E_x$ and electron density $n_e$ at $t=-1T_0$, respectively. It is evident that strong SPWs can still be efficiently excited at each front edge [\ref{fig:fig2}(e)] and electron nanobunches are periodically peeled off at each edge [\ref{fig:fig2}(h)]. The amplitude of the $E_x$ field locating at the central tape is almost comparable to that with a single tape since it is mainly decided by the amplitude of the laser field. However, here when the SPWs excited at different edges propagate forward along each target lateral surface, they will interfere with each other, as shown in Fig. \ref{fig:fig2}(e). This may interrupt the electron acceleration, since electrons may jump from the acceleration phase to the deceleration phase, leading to less energetic electrons. The corresponding energy spectrum at $t=29T_0$ is shown by the blue line in Fig. \ref{fig:fig7}(a), where one can see that the maximum electron energy is only about 150 MeV and the effective temperature is about 13.3 MeV. Both are much lower than those obtained from the single tape case, see the red line in Fig. \ref{fig:fig7}(a). The number of electrons with $\gamma>10$ is about 109.3 nC, 2.64 times higher than that from the single tape case. Though the electron number is much higher compared to the single tape (red line), it does not bring the increase of the hot electron density since electrons are extracted from more edges.  

When these electrons are injected into the vacuum, a strong longitudinally bunching field can still be induced, see Figs. \ref{fig:fig3}(b) and \ref{fig:fig4}(b). Since here the gap between the parallel tapes is relatively small and electrons diverge slowly during the propagation, the peak values of the $E_x$ field behind the rear edges are comparable while the center one is still slightly stronger. Therefore, protons from different rear edges experience similar acceleration process [see Fig. \ref{fig:fig3}(h)] and are accelerated to comparable energies from each tape. In Figs. \ref{fig:fig3}(h) and \ref{fig:fig4}(h), one can clearly see the density peaks at the frontmost of the proton density distributions, though the peak density is lower than that from a single tape, see Figs. \ref{fig:fig5}(a) and \ref{fig:fig5}(b). 

The proton energy spectrum at $t=54T_0$ is shown by the blue line in Fig. \ref{fig:fig7}(b), where the maximum and peak energies are 45.2 MeV and 41.1 MeV, respectively. There are about $2.1\times10^9$ protons inside the peak (FWHM), about one order of magnitude higher than that from the single tape case since protons are accelerated from three edges and the energy spread is relatively larger in this case. The conversion efficiency from laser to protons is about 1$\%$, almost comparable to that from the standard TNSA mechanism by using femtosecond laser pulses \cite{Green2014}. This will be useful for practical applications such as neutron sources and plasma heating. Note that besides those highly collimated energetic protons, we observe some protons mainly expanding transversely, as shown in Figs. \ref{fig:fig3}(g) and \ref{fig:fig3}(h). In the parallel tape case, when those transversely expanding protons from different tapes traverse with each other, the proton density becomes relatively higher than the surrounding regions, as shown in Fig. \ref{fig:fig3}(h), but this density peak will disappear once they pass through each other. This also explains why we see a second peak at $t=49T_0$ in Fig. \ref{fig:fig5}(b). 

\subsection{Results from a cross tape target}
 
In the case of a cross tape target, since the center of this target situates in the planes of both $y=0$ and $z=0$, at $t=-1T_0$, we show the 2D cuts of the fields and electron density in $z=0.9\mu$m and $y=0.9\mu$m planes at the third and fourth columns of Fig. \ref{fig:fig2}, respectively. One can also clearly see the strong longitudinal field and periodically-distributed electron nanobunches. Since a $y$-polarized laser pulse is considered here, the distributions in the planes of $z=0.9\mu$m (the third column in Fig. \ref{fig:fig2}) and $y=0.9\mu$m (fourth column) are not identical. In the plane of $z=0.9\mu$m, the main results are similar to those with a single tape shown in the first column, except that the electron density and the amplitude of the $E_x$ field are slightly lower since the plane deviates from the center of the laser pulse and the horizontal tape may degenerate the excitation of the SPWs. Though there are no laser fields for extracting electron nanobunches from the horizontal tape, due to refluxing and self-generated fields, we can still obtain more electrons from this setup, but mainly low-energy ones. The electron energy spectrum is shown by the black line in Fig. \ref{fig:fig7}(a). There are 50.6 nC electrons with $\gamma>10$, while the number of electrons with energy below 25 MeV is about 1.5 times more than that from the single tape case.. The maximum energy is about 214 MeV and the effective temperature is 29.2 MeV. Both are slightly lower than those obtained with a single tape target. 

However, since the cross tape separates the space into four quarters,  the propagation of the laser pulse and electron nanobunches and the geometry of the self-generated fields are quite different from those observed in the single tape case. Especially when the electron nanobunches are injected into vacuum, along $z$-direction they are not focused by the transverse fields into the center any more. As shown in Fig. \ref{fig:fig4}(f), there are two populations of electrons above and below the  $z=0$ plane and each of them are focused independently. Therefore, though the electron nanobunches are still well-collimated [see Fig. \ref{fig:fig6}(c)], the electron density is relatively low, especially behind the cross center. We can see an interesting double peak profile in the proton density [Figs. \ref{fig:fig3}(i) and \ref{fig:fig4}(i)]. The peak density is much lower than the other two cases discussed before and so is the velocity, see Fig. \ref{fig:fig5}. The black line in Fig. \ref{fig:fig7}(b) represents the final proton energy spectrum. The maximum energy is only about 33.4 MeV. There are two peaks in the spectrum, i.e., one is at energy about 22.3 MeV and the other is 31.4 MeV, corresponding to the two different density peaks shown in Figs. \ref{fig:fig3}(i) and \ref{fig:fig4}(i). There are about $2.9\times10^9$ protons within the lower peak (FWHM) and the conversion efficiency from laser to protons is about $0.46\%$, comparable to the single tape case. We mention that for this cross tape target, it would be better to use a circularly-polarized laser pulse where more electrons can be peeled off from both the horizontal and vertical tapes, leading to more efficient ion acceleration. Moreover, radially-polarized laser pulses would also be an option \cite{KARMAKAR2007,Wen2019,Naim2020,Li2012,Cao2021}.

\begin{figure*}[h]
	\includegraphics[width=14cm]{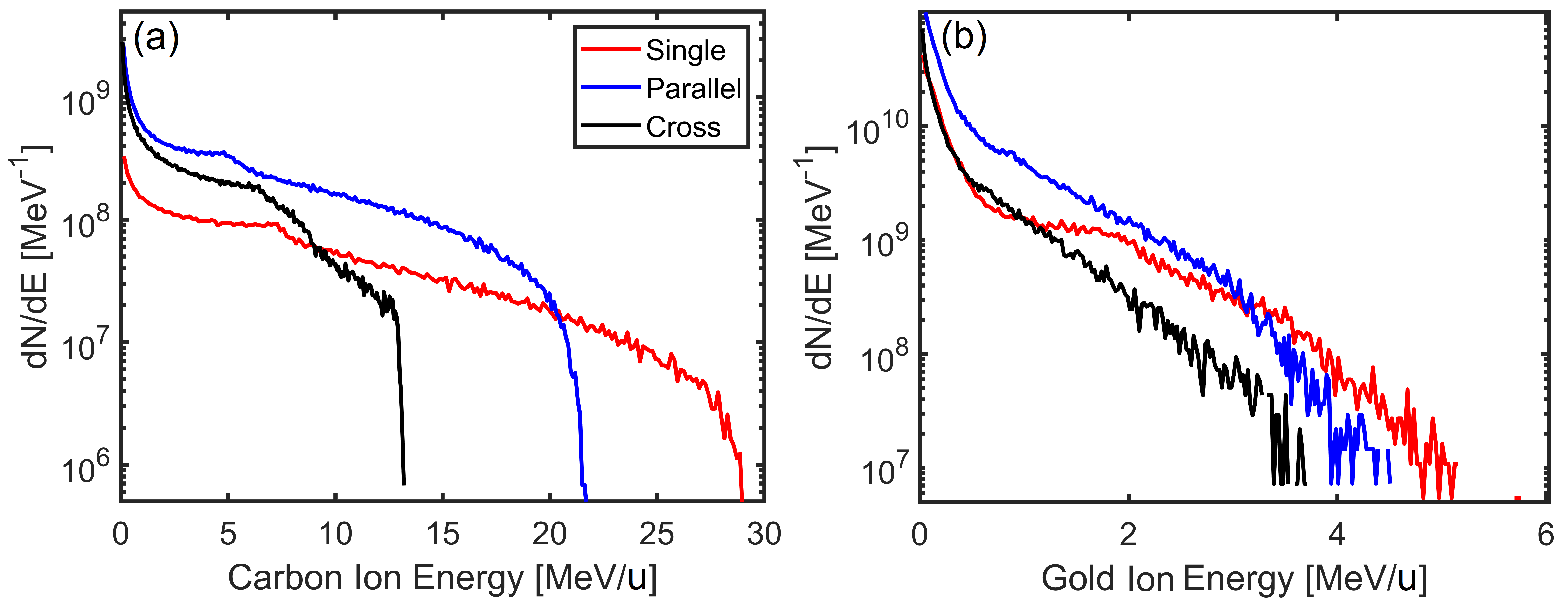} \caption{Heavy ion acceleration from the peeler scheme: (a) Energy spectra of carbon ions from the single (red), parallel (blue) and cross (black) tape target at $t=54T_0$. Here carbon ions are selected within $10^\circ$. (b) The corresponding spectra of gold ions at $t=54T_0$. Note that for gold ions, most of them are accelerated transversely so we select them with conditions of $|\theta-90^\circ|<10^\circ$ and $p_y>0$.
	}
	\label{fig:fig8} 
\end{figure*}

\subsection{Heavy ion acceleration}

Besides protons, the tape target  has components of carbon ions $\rm C^{6+}$ and gold ions $\rm Au^{51+}$. They can also be accelerated by the quasistatic fields, though not as efficient as protons since they have smaller charge-to-mass ratios. The carbon ions are mainly accelerated forward, following the faster protons. Since the total charge of the carbon ions is six times larger than that of the protons, the charge density of electrons around the high-energy carbon ions is lower than that of the carbon ions. Therefore, the accelerating field for carbon ions is a debunching field, leading to an exponentially decaying energy spectrum. In Fig. \ref{fig:fig8}(a), the red, blue and black lines represent the final spectra obtained from the cases with a single, parallel and cross tape target, respectively. The maximum energies of carbon ions are 28.9 MeV/u, 21.7 MeV/u and 13.2 MeV/u, respectively. The conversion efficiency from the laser to carbon ions is about $1.9\%$, $3.6\%$ and $1.4\%$, respectively.

For gold ions, they are mainly accelerated along the $y$-direction, pulled out by the peeled-off electron nanobunches. Their spectra are correspondingly shown in Fig. \ref{fig:fig8}(b), where gold ions are selected with conditions of $|\theta-90^\circ|<10^\circ$ and $p_y>0$. They also show an exponentially decaying profile. The maximum energies from the single tape is still the highest, about 5.8 MeV/u (i.e., 1.14 GeV per atom), see the red line. Here the difference of the maximum energies obtained from these three setups is not large. This is because the origin of the accelerating field of the gold ions is similar to the capacitor-like electric field \cite{Shen2019} which is mainly determined by the density of the peeled-off electrons, and has little relation to the electron temperature, divergence and propagation. The conversion efficiency from the laser to gold ions is about $7.8\%$, $17\%$ and $5.6\%$ for the single, parallel and cross tape cases, respectively. Then, we know that the corresponding total conversion efficiency from the laser to ions (including protons, carbon and gold ions) is about $10.2\%$, $21.6\%$ and $7.5\%$. Note that in the case with a parallel tape target, to be detected in experiment, gold ions from the center tape traverse the outer tape. However, since the thickness of the tape target in $y$-direction is only 0.75$\lambda$, the energy loss is negligible \cite{Gibbon2005}.

\begin{figure*}[h]
	\includegraphics[width=14cm]{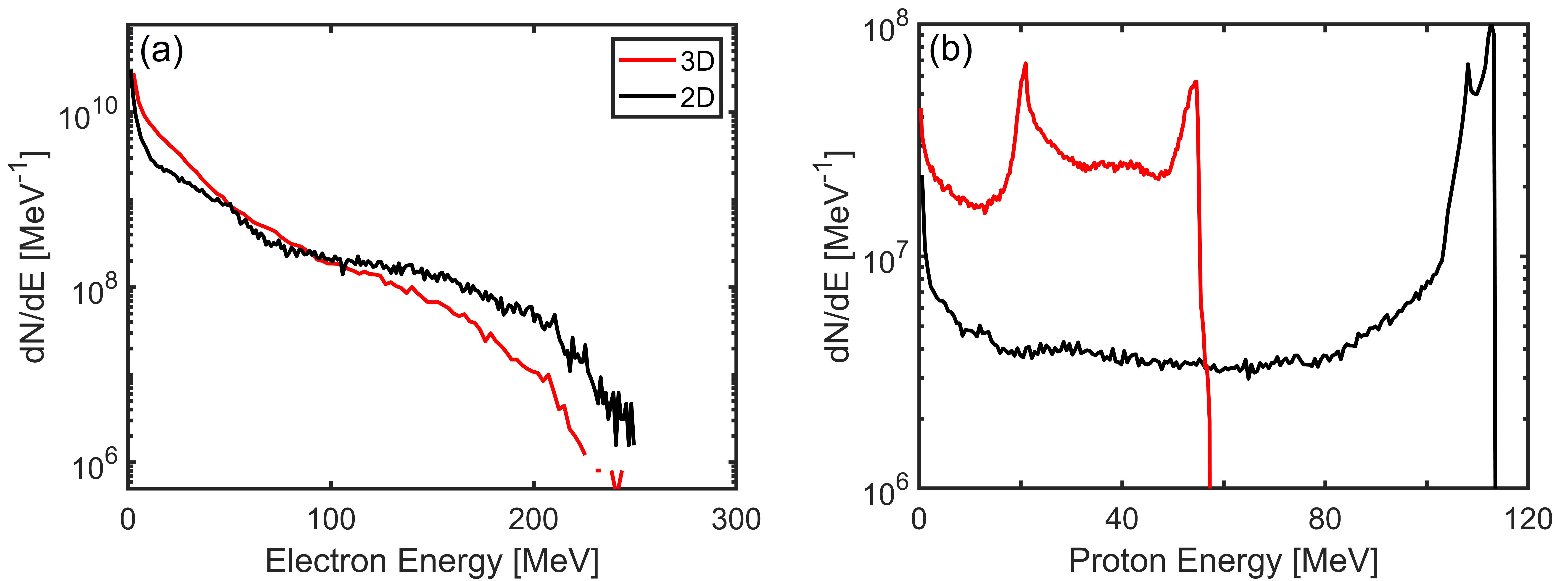} \caption{Comparison of 2D (black) and 3D (red) simulation results from the single tape case: (a) Electron energy spectra at $t=29T_0$ and (b) proton energy spectra at $t=54T_0$. Note that here we use a prefactor of $\pi r_L [\mu m]/2$ to convert the particle number obtained from 2D simulations.
	}
	\label{fig:fig9} 
\end{figure*}

\section{Discussion}

To quantitatively compare with experimental results and also investigate the full dynamics, 3D PIC simulations are necessary, but they usually consume large amount of computational time. 
Therefore, lots of numerical studies are still based on 2D simulations. Here we further investigate the dimensional effects in our scheme and only focus on the single tape case.

In 2D simulations, the laser and plasma parameters remain the same. The sizes of the simulation box in $x$- and $y$-directions also remain unchanged, but we use better resolutions in both directions. Now there are 5600$\times$1920 cells along $x\times y$ directions. The macroparticles in each cell for electrons, Au$^{51+}$, C$^{6+}$ and H$^{+}$ are correspondingly increased to 100, 50, 200 and 800. The main physical processes remain almost the same so we only compare the obtained energy spectra of electrons and protons here.

Figure \ref{fig:fig9} shows the comparison between 2D (black) and 3D (red) simulations. In Fig. \ref{fig:fig9}(a), it is evident that the maximum electron energies are almost the same since these electrons are well-collimated and in 3D, their motion is confined in the $z=0$ plane. The electron number and temperature are not the same. This can be attributed to the nonuniform distributions of electrons and fields along $z$-direction in 3D. Interestingly but not surprisingly that the maximum proton energy in 3D is about two times lower than that in 2D. Similar results have been discussed in the TNSA mechanism \cite{Xiao2018,Sgattoni2012}. The reason can be attributed to that in 3D, the electron density decreases faster during the propagation since electrons have one extra dimension to diverge \cite{Xiao2018,Sgattoni2012}. A quantitative explanation will be future work. 

In our simulations, a thick CH layer is considered to mimic the contaminant to avoid protons being accelerated as test particles. In experiments, the thickness of the contaminant is about a few nm, but the density can be much higher \cite{Allen2004,Lecz2020}. The available proton number can be $10^{11}$ or even higher \cite{Allen2004,Badziak2018,Macchi2013,Daido2012}. In our scheme, though protons are only located at the edge of one or multi tape targets, the available proton number within the laser focal spot size is actually not very small compared to the other known setups. One can easily estimate the ratio of the available proton number in the single tape regime ($N_p$) to the other known setup ($N_p^s$) as $\eta=N_p/N_p^s=4l_y/\pi d_L$. For a tightly-focused laser pulse with $d_L\sim4\lambda$ \cite{Wagner2016,Bin2015} and $l_y=0.75\lambda$ considered in our simulations, $\eta\approx0.24$. Therefore, it is possible to obtain $\sim10^9$ protons in experiment even with the single tape target.

Moreover, in our previous work \cite{Shen2021b}, we have demonstrated the effectiveness of this scheme by considering some realistic parameters, including CH preplasma wrapped around the target, oblique incidence and higher electron density. It indicates that our scheme can be validated via a plastic tape target. Only when the laser intensity is ultra high or the pulse duration is very long, one needs to consider the effects induced by the target transverse expansion as aforementioned. 
The investigated preplasma scale can be realized with the state-of-the-art laser systems \cite{Henig2009,Danson2015}. 
If the laser pulse is incident on the tape at a small angle such as 2$^\circ$, only the peak energy drops a little bit and the other parameters stay almost the same, see blue line in Fig. 6(a) of Ref. \cite{Shen2021b}. If the incidence angle is 10$^\circ$, the peak energy is reduced by about 20$\%$ while the energy spread stays extremely low, see the Supplementary Note 2 and Fig. S1 in Ref. \cite{Shen2021b}.
Therefore, we think that it is quite promising to validate this peeler scheme, at least observe the quasimonoenergetic structure, in current laser systems, even without very good pointing stability. 

\section{Conclusion}

In conclusion, we further discuss the electron and ion acceleration in laser-plasma peeler scheme by considering a tabletop 200 TW-class femtosecond laser. Two more variants, i.e., a parallel tape and a cross tape target, are also investigated. Our 3D simulations demonstrate that quasimonoenergetic proton beams can be obtained in all the three setups, though the detailed beam parameters (e.g. maximum energy, peak energy, energy spread and flux) are different. 
Compared to the two variants, the single tape target is favorable for producing proton beams with higher peak energy and narrower spectrum. This is because the laser energy concentrates on a single tape and electrons can stay in phase with SPWs for longer time and are strongly focused by the transverse fields, leading to a stronger and longer-lasting bunching field for protons. Moreover, we mention that the results shown here can be further optimized. Even for the same laser parameters, one can further optimize the results by adjusting the length ($x$-direction), thickness ($y$-direction) or material of the tape target. The effects of dimensionality on PIC simulations of this peeler scheme are also discussed.

\section*{Data availability statement}
The data that support the findings of this study are available upon request from the authors.

\ack
This work is supported by the DFG (project PU 213/9). 
The authors gratefully acknowledge the Gauss Centre for Supercomputing e.V. (www.gauss-centre.eu) for funding this project by providing computing time through the John von Neumann Institute for Computing (NIC) on the GCS Supercomputer JUWELS at Jülich Supercomputing Centre (JSC). X.F.S. gratefully acknowledges support by the Alexander
von Humboldt Foundation, as well as acknowledges helpful discussions with L. Reichwein at HHU and Y. L. Yao at PKU.

\end{document}